\documentclass[grl]{agutex}

 \usepackage{lineno}
  \usepackage{graphicx}
  \setkeys{Gin}{draft=false}
\usepackage{color}

\authorrunninghead{EDBERG ET AL.}

\titlerunninghead{PLASMA AROUND COMET 67P}

\authoraddr{N. J. T. Edberg, M. Andr{\'e}, A. I. Eriksson, R. Gill, E. Johansson, F. Johansson, E. Odelstad, E. Vigren,  J.-E. Wahlund,
Swedish Institute of Space Physics, Uppsala, Box 537, SE-75121, Sweden}
\authoraddr{P. Henri, J.-P. Lebreton, Laboratoire de Physique et Chimie de l'Environnement et de l'Espace, OrlŽans Cedex 2, France}
\authoraddr{H. Nilsson, G. Stenberg Wieser, Swedish Institute of Space Physics, Box 812, 981 28 Kiruna, Sweden}
\authoraddr{C. Carr, E. Cupido, Imperial College London, Exhibition Road, London SW7 2AZ, United Kingdom}
\authoraddr{S. Gasc, M. Rubin, Physikalisches Institut, University of Bern, Sidlerstrasse 5, CH-3012 Bern, Switzerland}
\authoraddr{R. Goldstein, K. Mandt, Southwest Research Institute, 6220 Culebra Rd., San Antonio, TX 78238, USA}
\authoraddr{C. Carr, E. Cupido, Imperial College London, Exhibition Road, London SW7 2AZ, UK}
\authoraddr{K.-H. Glassmeier, I. Richter, C. Koenders, TU - Braunschweig, Institute for Geophysics and extraterrestrial Physics, Mendelssohnstr. 3, D-38106 Braunschweig, Germany  }
\authoraddr{K. Szego, Z. Nemeth, Wigner Research Centre for Physics, 1121 Konkoly Thege street 29-33, Budapest, Hungary}
\authoraddr{M. Volwerk, Space Research Institute, Austrian Academy of Sciences, Schmiedlstrasse 6, 8042 Graz, Austria}

\begin{document}

%
%
\title{Spatial distribution of low-energy plasma around comet 67P/CG from Rosetta measurements}

%
%
\authors{N. J. T. Edberg, \altaffilmark{1}
A. I. Eriksson, \altaffilmark{1}
E. Odelstad, \altaffilmark{1,2}
P. Henri, \altaffilmark{3}
J.-P. Lebreton, \altaffilmark{3}
S. Gasc, \altaffilmark{4}
M. Rubin, \altaffilmark{4}
M. Andr{\'e}, \altaffilmark{1}
R. Gill, \altaffilmark{1}
E. P. G. Johansson, \altaffilmark{1}
F. Johansson, \altaffilmark{1}
E. Vigren, \altaffilmark{1}
J. E. Wahlund, \altaffilmark{1}
C. M. Carr, \altaffilmark{5}
E. Cupido, \altaffilmark{5}
K.-H. Glassmeier, \altaffilmark{6}
R. Goldstein,\altaffilmark{7}
C. Koenders, \altaffilmark{6}
K. Mandt,\altaffilmark{7}
Z. Nemeth, \altaffilmark{8}
H. Nilsson, \altaffilmark{9}
I. Richter, \altaffilmark{6}
G. Stenberg Wieser, \altaffilmark{9}
K. Szego, \altaffilmark{8}
M. Volwerk, \altaffilmark{10}
}

\altaffiltext{1}{Swedish Institute of Space Physics, Uppsala, Sweden (ne@irfu.se)}
\altaffiltext{2}{Department of Physics and Astronomy, Uppsala University, Box516, SE-75120 Uppsala, Sweden}
\altaffiltext{3}{Laboratoire de Physique et Chimie de l'Environnement et de l'Espace, Orleans, France}
\altaffiltext{4}{Physikalisches Institut, University of Bern, Bern, Switzerland}
\altaffiltext{5}{Space and Atmospheric Physics Group, Imperial College London, UK} 
\altaffiltext{6}{TU - Braunschweig, Institute for Geophysics and extraterrestrial Physics, Braunschweig, Germany}
\altaffiltext{7}{Southwest Research Institute, San Antonio, USA}
\altaffiltext{8}{Wigner Research Centre for Physics, Budapest, Hungary}
\altaffiltext{9}{Swedish Institute of Space Physics, Kiruna, Sweden}
\altaffiltext{10}{Space Research Institute, Austrian Academy of Sciences, Austria}









%
%


\begin{abstract}
We use measurements from the Rosetta plasma consortium (RPC) Langmuir probe (LAP) and mutual impedance probe (MIP) to study the spatial distribution of low-energy plasma in the near-nucleus coma of comet 67P/Churyumov-Gerasimenko. The spatial distribution is highly structured with the highest density in the summer hemisphere and above the region connecting the two main lobes of the comet, i.e. the neck region. There is a clear correlation with the neutral density and the plasma to neutral density ratio is found to be $\sim$1-2$\cdot10^{-6}$, at a cometocentric distance of 10 km and at 3.1 AU from the sun. A clear 6.2 h modulation of the plasma is seen as the neck is exposed twice per rotation. The electron density of the collisonless plasma within 260 km from the nucleus falls of with radial distance as $\sim$$1/r$. The spatial structure indicates that local ionization of neutral gas is the dominant source of low-energy plasma around the comet. 
\end{abstract}

%
%

%

\begin{article}

%
%

\section{Introduction}
Comets originate from the early formation of the solar system and consist to a large extent of volatile material, such as ice from H$_2$O, CO and CO$_2$. The ice will eventually begin to sublimate due to heating by sunlight and produce gas, which expands around the comet and leads to the formation of a coma. The coma subsequently gets partly ionized and an ionosphere is formed in the inner coma. The extreme ultraviolet radiation (EUV) is the primary source of ionization but particle impact and charge exchange processes also contribute \citep{cravens1987, nilsson2015}.
 
Comet 67P/Churyumov-Gerasimenko (hereafter called 67P) currently orbits the sun with a period of 6.44 years in an elliptical orbit with perihelion at 1.24 AU and aphelion at 5.68 AU. When orbiting far out in the solar system comets in general are relatively inactive, but when approaching the sun the heating and outgassing increase. Since 6 August 2014, starting at 3.6 AU from the Sun, the European Space Agency's Rosetta mission has followed 67P closely along its orbit. One of the objectives of Rosetta is to monitor the plasma environment in the cometary coma \citep{glassmeier2007}. 

Early signs of a cometary plasma environment around 67P, through the detection of water ions, were observed already at 3.6 AU from the sun $\sim$100 km from the comet \textcolor{black}{\citep{nilsson2015}. Cometary ions were being picked up by the solar wind motional electric field and accelerated back toward the comet and Rosetta \citep{nilsson2015,goldstein2015}. The total water production rate at 3.6 AU} was measured to be about $4\cdot10^{25}$ molecules s$^{-1}$ \citep{gulkis2015}. \textcolor{black}{The photoionization frequency of H$_2$O at 1 AU is about $3-8\cdot10^{-7}$ s$^{-1}$ depending on solar conditions, and should fall off as the square of the heliocentric distance  \citep{huebner2015}}. Therefore only a small fraction of the cometary molecules gets ionized in the vicinity of the comet. The principle of quasi-neutrality should ensure that the ion and electron densities are the same, as long as there is relatively little dust present to which electrons can attach. When significant amounts of dust is present, the free electron density can decrease significantly in comparison to the ion density.

Besides water, CO and CO$_2$ are also abundant in the comet nucleus and sublimates to contribute to the coma gas composition. \citep{hassig2015} reported from Rosetta measurements that the composition of the coma varies with the rotation phase of the comet and which side is facing the sun. Different illumination conditions and potentially thermal processing of the surface leads to a strong variation in the relative abundances of the major species measured in the coma. CO$_2$, which is both heavier than water and ionizes three times faster, is more dominant in the southern latitudes \citep{hassig2015}. The ionization of these molecules leads primarily to H$_2$O$^+$, OH$^+$, H$^+$, CO$^+$,  \textcolor{black}{CO$^{2+}$,} O$^+$ and C$^+$. Closer to perihelion, ion neutral reactions will rapidly synthesize H$_3$O$^+$ from these primary ions and H$_3$O$^+$ can in turn proton transfer to e.g. ammonia and methanol \citep{vigren2013b}. 

A partly ionized and collisionless coma of an active comet will interact with the solar wind and many regions and plasma boundaries will form as a consequence; see e.g. \citet{koenders2013} for a simulation of the comet-solar wind interaction. The innermost boundary, which Rosetta could possibly see at these distances ($>$3 AU), is the contact surface at which the outward directed ion-neutral drag force is balanced by the inward directed magnetic gradient pressure force. 

The plasma environments of comets have previously only been measured at relatively large distances from the nucleus and mainly for active comets close to the sun. Never before has it been possible to measure the structure of the plasma environment of a weakly outgassing comet at distances $<$ 10 km from the comet nucleus. With Rosetta we are now able to explore this region and in this paper we will \textcolor{black}{present measurements of the spatial distribution of the plasma around comet 67P, when 2.2-3.2 AU from the Sun and when $\sim$8-260 km from the comet centre}.

\section{Instruments and measurements}
The Rosetta spacecraft \citep{glassmeier2007} carries a suite of five instruments, forming the Rosetta plasma consortium (RPC) \citep{carr2007}, to measure the plasma properties as well as electric and magnetic fields around comet 67P. Of particular interest here is the Langmuir probe instrument (LAP) \citep{eriksson2007} and the mutual impedance probe (MIP) \citep{trotignon2007}. Rosetta also carries the Rosetta orbiter spectrometer for ion and neutral analysis (ROSINA) instrument, including the Comet pressure sensor (COPS), which measures the density and dynamic pressure of the neutral gas around the comet \citep{balsiger2007}. In this paper we will use ion and electron density estimates from LAP, electron density estimates from MIP \textcolor{black}{and} neutral gas density measurements from ROSINA. 

LAP consists of two spherical Langmuir probes (LAP1 and LAP2), which are fastened on stubs and mounted on booms, 2.2 m and 1.6 m long, respectively, from hinge to probe. Their orientation is such that at least one probe is always in sunlight (so far mostly LAP1) and at least one in the radial flow from the comet (so far mostly both probes). 

The MIP instrument consists of two receiving and two transmitting electrodes, mounted on the same boom as LAP1. MIP can retrieve plasma parameters only when the ratio of the emitter-receiver baseline length to the Debye length is large enough. For low densities, MIP makes use of LAP2 as transmitter, at some 4 m distance from the MIP receivers \citep{trotignon2007}. In this so-called long Debye length (LDL) mode, MIP can record densities down to some tens of cm$^{-3}$, depending on the electron temperature T$_e$.

Each Langmuir probe can operate independently from the other and, intermittently, LAP2 is used in the MIP LDL mode. In this paper we will only use data from LAP1 when in `sweep' mode, to obtain the ion and electron density, and from MIP when used in LDL mode to obtain electron density. In the LAP sweep mode the probe sweeps the bias potential applied to the probe from a minimum of -32 V to a maximum of +32 V and collects either the ion or the electron current, depending on the sign of the potential. Such a sweep is carried out with a cadence of any multiple of 32 s. In the interval used in this paper the cadence of the LAP sweep measurements is 96 s or 160 s, depending on telemetry available. 

From the measured current-voltage curve the electron density and temperature, ion density and speed, mean ion mass as well as the spacecraft potential can be extracted. Due to the usually high negative spacecraft potential \textcolor{black}{($\sim$ -10 V)} observed during the early phase of the mission a substantial fraction of the electrons are out of reach and electron plasma parameters are therefore challenging to derive. On the ion side (negative probe bias voltage) the current is less dependent on the spacecraft potential. In orbit motion limited theory \citep{fahleson1974} the ion and electron densities from the sweeps are given by

\begin{equation}
 N_i = \sqrt{ \frac{a b m_i}{2 \pi^2 r^4 e^3} },
\end{equation} 

and 

\begin{equation}
 N_e = \frac{dI_{e}}{dV}\frac{T_e}{eA}\sqrt{\frac{2\pi m_e}{k_bT_e}},
\end{equation} 

where $N_i$ is the ion density, $a$ is the current value at $V=0$ (when the photoelectron current of $\sim$ -8.5 nA, determined when the probe moves from sunlight to shadow at a heliocentric distance of 3.2 AU, has been subtracted), $b$ is the slope of the curve on the ion side, $m_i$ the average ion mass (the main ion is assumed to be H$_2$O$^+$ with mass $m_i$=18 amu), $r=0.025$ m is the probe radius, $e$ the elementary charge, $N_e$ is the electron density, $V$ the bias voltage with the spacecraft potential subtracted, $dI_e/dV$ is the slope of the current on the electron side, $T_e$ is the electron temperature, $A=4\pi r^2$ is the probe area, $m_e$ is the electron mass, and $k_B$ is Boltzmann's constant. 

Error estimates show that the main uncertainty in the density estimate is caused by the offset in the current due the uncertain photoelectron current. An uncertainty of 0.5 nA in the photocurrent translates to an error in $N_{i}$ of up to 50$\%$. Regular calibrations are performed onboard to identify and remove any slope and offset error, caused by the electronics, from the current measured. 

The coordinate systems used to describe the Rosetta measurements are the Comet solar orbital (CSO) system and the CK system (which is almost identical to the Cheops system). In the CSO system the $X_\mathrm{CSO}$-axis is directed toward the sun, the $Z_\mathrm{CSO}$-axis is perpendicular to $X_\mathrm{CSO}$ and projected onto the vector of the orbital plane and the $Y_\mathrm{CSO}$-axis completes the right-handed system and is directed, approximately, opposite to the comet's orbital velocity vector. In the comet-fixed coordinate system (CK system) the $X_\mathrm{fixed}$-axis is pointing outward from the mass-centre toward the minor lobe of the comet, the $Z_\mathrm{fixed}$-axis is parallel to the comet spin-axis and the $Y_\mathrm{fixed}$-axis completes the right-handed system. 

\section{Observations}
\begin{figure}
  \noindent\includegraphics[width=16cm]{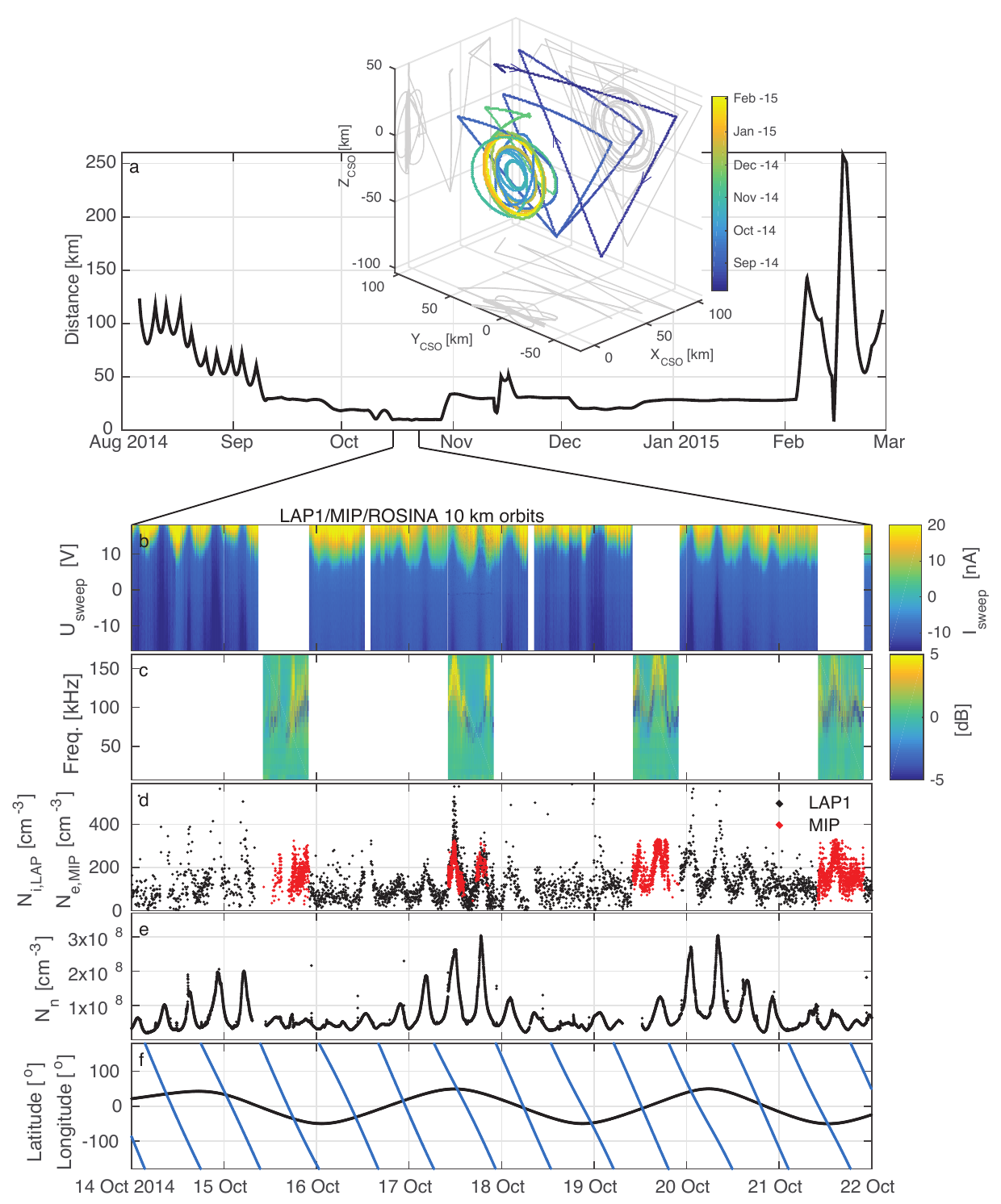}
 \caption{Time series of Rosetta RPC-LAP/MIP data from the bound orbits at 10 km distance. The individual panels show (a) the cometocentric distance of Rosetta, with the inset showing the trajectory of Rosetta around the comet in CSO coordinates with time color-coded along the track, (b) sweep data from LAP1 where the bias voltage is shown swept from -18V to +18 V and the collected current is color-coded, (c) active spectrogram from MIP (d) derived ion density from the LAP1 sweeps (black) and electron density measured by MIP (red), (e) ROSINA/COPS neutral density and (f) latitude (black) and longitude (blue).}
 \label{fig:time}
 \end{figure}

The RPC instruments began operating regularly on 9 May 2014 at a distance of $1.8 \cdot 10^6$ km from the comet and at a heliocentric distance of 4.1 AU. During the following 3 months Rosetta decreased its distance to the comet to reach 120 km on 6 Aug \textcolor{black}{2014}. Figure \ref{fig:time}a illustrates how Rosetta, from 6 August 2014 until 8 September, approached the comet further by following triangular paths at decreasing distances from 120 km to 30 km. Early signatures of cometary plasma started to appear in LAP data in early August. Not until the triangular/pyramid paths began (heliocentric distance of about 3.6 AU) did the cometary plasma become clearly observable and started to dominate over the spacecraft generated photoelectron cloud. The early signatures were mainly identified through a periodically changing spacecraft potential, due to an increase in low energy electrons, as well as an increase in electron and ion currents to the probe. The signatures appeared with a periodicity of 6.2 hours, i.e. half the comet rotation period. No bow shock or other type of plasma boundary arising from the solar wind interaction was observed during the approach phase.

On 8 September bound orbits close to the terminator plane started, at decreasing cometocentric distances of 30 km, 20 km and 10 km.  
In Figure \ref{fig:time} (lower part) we show a time series of Rosetta LAP1, MIP and ROSINA/COPS data as well as Rosetta's cometary longitude and latitude from the interval when at a cometocentric distance of $\sim$10 km. During the two weeks in 10 km orbits, Rosetta's heliocentric distance decreased from 3.17 AU to 3.09 AU. The comet spin axis was about 48 degrees from the comet-sun line such that the northern hemisphere experienced summer at this time. LAP1 was at this time in sunlight and in the undisturbed radial flow from the comet. In Figure \ref{fig:time}b the LAP1 sweeps go from at least -18V to +18V throughout the interval. 

Intermittently, the MIP instrument was operating in LDL mode to provide active spectrograms, shown in Figure \ref{fig:time}c). This data is shown after computation of an estimate of the MIP response in vacuum, used to obtain a normalized plasma-to-vacuum mutual impedance. The electron density is estimated from these spectrograms, from the position of the plasma frequency, assuming Maxwellian electrons in a locally homogeneous plasma and taking into account the influence of the nearby conducting spacecraft structure \citep{geiswiller2001}. The electron density is plotted in Figure \ref{fig:time}d, together with the LAP1 sweep-derived ion densities. The electron and ion densities are found to typically be of the order of 100 cm$^{-3}$, at 10 km distance to the comet when at $\sim$ 3.1 AU from the sun, but the ion density can occasionally be a factor of 5-10 higher. Note that in this interval the MIP LDL mode density saturates above about 350 cm$^{-3}$ due to the frequency limit of the instrument.

The gas formed from sublimation is flowing radially outward with a velocity of the order of 700 m s$^{-1}$ \citep{gulkis2015} and is being partly ionized, through predominantly photoionization. The density of the neutral gas, measured by the ROSINA/COPS, instrument are plotted in Figure \ref{fig:time}e. Our measurements at 10 km indicate that the ratio between plasma density and neutral density is typically 1-2${\cdot}10^{-6}$, when comparing panel d and e of Figure \ref{fig:time}. The peaks in ion, electron and neutral density (coinciding with dips in spacecraft potential) occur when Rosetta is above the neck-area, i.e. in between the two main lobes of comet 67P at longitudes of approximately $+60^{\circ}$ and $-120^{\circ}$ and creates a 6.2-hour periodicity to the data. This has been seen since arrival at the comet in early August.

Photoionization of H$_2$O, CO and CO$_2$ produces electrons with a temperature of about 15 eV \citep[e.g.][]{cravens1987}. These would cool due to collisions if the neutral density was high enough, but the LAP measurements indicate that they were not yet significantly cooling when at these 10 km orbits, and the neutral coma was, hence, relatively tenuous at this time. Consequently, the spacecraft is charged to a negative potential from the impacting flux of these hot electrons. A large fraction of the electron distribution in this interval is sometimes out of reach of the LAP instrument due to the very negatively charged spacecraft. Hence, we do not present sweep derived electron densities from this interval when at 10 km but rather only electron densities from MIP, which should be more reliable when in the LDL mode. 

Besides the 6.2-hour variation, there is also a clear trend of higher peak densities when at northern latitudes. The northern `hemisphere' was experiencing summer at this time and was therefore more heated by sunlight. There is generally a strong correlation between the plasma and neutral density around the comet and both peak in the northern hemisphere. 

\begin{figure}
 \noindent\includegraphics[width=17cm]{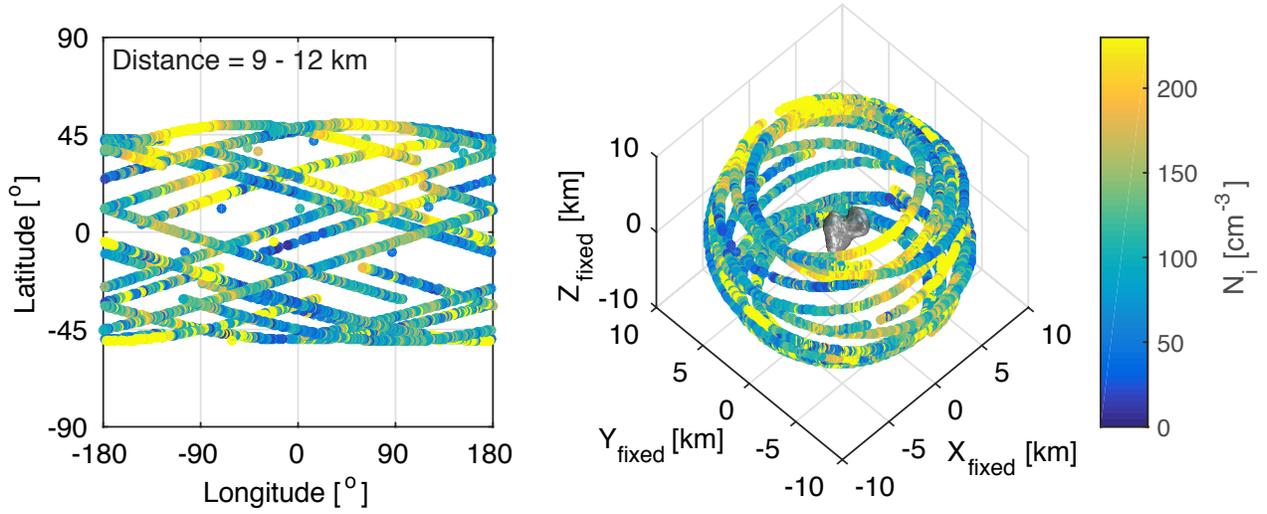}
 \caption{Ion density around comet 67P from the 10 km orbits. The LAP1 sweep derived ion density is color-coded on a longitude-latitude map (left) as well as in a 3D-view in comet-fixed coordinates. A shape-model of the comet from OSIRIS images is included in the 3D-view. The highest plasma density is observed over the neck-region of the comet at roughly $+60^{\circ}$ and $-120^{\circ} $ longitude, and in the summer hemisphere ($+Z_\mathrm{fixed}$).}
 \label{fig:ionmap}
 \end{figure}

To illustrate the structured spatial distribution of plasma around comet 67P, we show in Figure \ref{fig:ionmap} the ion density colour-coded in both a longitude-latitude map as well as in a 3D plot in comet-fixed coordinates using the measurements taken during the 10 km bound orbits. In the longitude-latitude map the neck-area is located at about $+60^{\circ}$ and $-120^{\circ}$ longitude. Two clear bands of higher densities (bright yellow) appear in this region for most latitudes covered by Rosetta, although the trend is clearer in the northern hemisphere. A strongly isotropic spatial distribution of the low-energy plasma is present in the near-nucleus environment.

The 3D-view plot in the right panel illustrates more clearly how the plasma is distributed around the comet. This plot has the comet in the centre and the density measured by LAP1 colour-coded along the trajectory of Rosetta. In the general direction toward the sun and where the neck is exposed ($\pm Y_\mathrm{fixed}$ and $+Z_\mathrm{fixed}$ direction) the plasma density (as well as the neutral density) are higher compared to elsewhere around the comet. 

These maps are more or less reproducible for the 30 km orbits earlier in September 2014, but more clearly for the 30 km orbits in late November 2014  and early January 2015, with the same general pattern emerging of a higher-density region located above the summer neck-area. When at 30 km in September, at 3.3 AU, the 6.2-hour variation is still clearly visible in the sweep data as well as in the MIP electron density measurements. The spatially structured plasma environment naturally suggests that local ionization dominates over solar wind plasma or picked-up comet plasma accelerated by the solar wind convective electric field, which would not have been spatially structured around the comet. The neutral gas is tenuous enough to be considered non-collisional, shortly after it leaves the comet nucleus.

\begin{figure}
 \noindent\includegraphics[width=12cm]{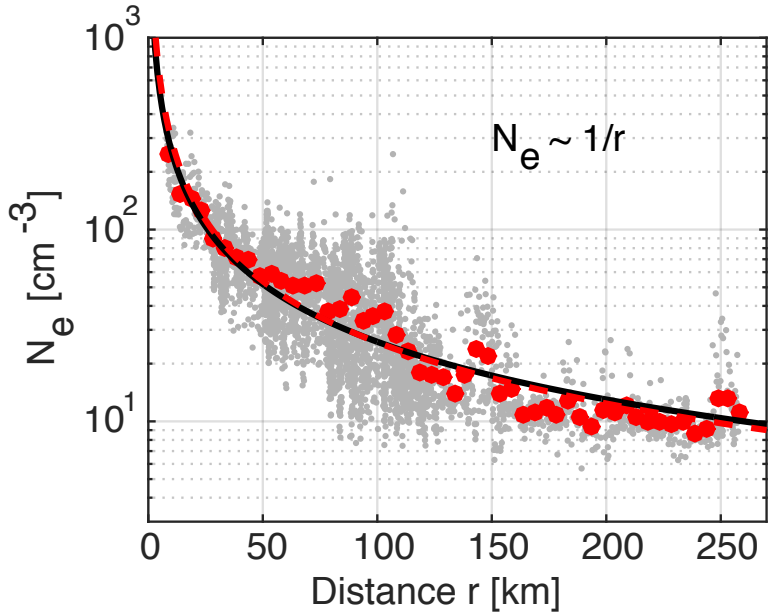}
 \caption{Electron density as a function of cometocentric distance (grey dots) and mean values over 5 km distance intervals (red dots). The red dashed line is a least-squares power-law fit to the mean values and the black solid line shows a $1/r$ dependence. The data is gathered from 4 - 28 Feb \textcolor{black}{2015}.}
 \label{fig:altprofile}
 \end{figure}

Next, we study how the density varies with distance from the comet. Figure \ref{fig:altprofile} shows an altitude profile of the LAP1 sweep electron density with data from the first two flybys from 4 - 28 February \textcolor{black}{2015} (see Figure \ref{fig:time}a). During these flybys Rosetta moved out to a maximum of 260 km from the nucleus and stayed at northern latitudes the entire time. When moving away from the comet, the spacecraft potential changes toward less negative and, eventually, a few volts positive such that in this interval the electron density measurement is more reliable than it was during the 10 km orbits. The grey dots in the scatter-plot show all the data but with intervals of spacecraft slews and intervals with LAP1 in shadow excluded. Mean values of the electron density (red dots) are computed in altitude intervals of 5 km. A $1/r$ dependence is indicated by a black solid line and it agrees very well with the data. A power-law function least-squares fitted to the median values gives the altitude dependence to vary as $1/r^{1.06}$. 

Following \citet{haser1957} the distribution with distance $r$ of the neutral density in a cometary coma, which is originating from a spherical source of radius $R_0$ and expanding radially outward and whose only loss source is an exponential decay with distance, can be described as

\begin{equation}
N_{n}(r) = n_{0}\Big(\frac{R_0}{r}\Big)^2 e^{\frac{r-R_0}{L}},
\end{equation}

where $n_0$ is the density at $r=R_0$. $L$ is the characteristic length scale of exponential loss of neutrals through ionization (assuming equal probability of ionization at all times, i.e. an optically thin coma). The distribution of the electron density formed from the exponentially decaying neutrals at a distance $r$ is then

\begin{equation}
N_{e}(r) = n_{0}\Big(\frac{R_0}{r}\Big)^2 \Big[1- e^{\frac{r-R_0}{L}}\Big] \sim \frac{\frac{r-R_0}{L}}{r^2} \sim \frac{1}{r}, 
\end{equation}

where a Taylor expansion is used in the second step. No recombination or other loss of electrons are assumed in this model. Hence, at large distances from the source ($R_0 \ll r \ll L$ ) the electron density should fall of as $1/r$, \textcolor{black}{which is what we observe in Figure \ref{fig:altprofile}}. 

\section{Summary and Discussion}
Plasma of cometary origin dominates the plasma environment around comet 67P already at 3.3 AU, when within 30 km ($\ll$ ion gyro radii) from the nucleus. The highest plasma density is observed above the neck area, i.e. the area in between the two main lobes of the comet, which is exposed to Rosetta twice per rotation, and in the northern summer latitudes. This does not necessarily mean that the neutral gas originate from the neck area on the comet. It could still be an effect of focusing or enhanced illumination of the irregular shape of the nucleus that causes these localised high densities. Nevertheless, the spatial distribution of the plasma is generally highly structured around the comet, indicating that local ionization of neutral gas is the main source of the plasma. This also indicates that the plasma is collisionless at this heliocentric distance. The LAP ion density and the MIP electron density are generally in very good agreement during the 10 km orbits. Both instruments do suffer from measurement uncertainties, which are primarily caused by a very negative spacecraft potential and an uncertainty in the photoelectron current estimate.

The electron density is found to fall off with distance as $1/r$, which is expected from ionization of a neutral gas expanding radially from the comet nucleus and when there is no significant recombination or other loss source for the plasma. \textcolor{black}{\citet{balsiger1986} reported on the plasma density altitude dependence from Giotto measurements at comet Halley when a fully developed coma had formed. They observed that the ion density decreased as $1/r$ inside the contact surface, where photochemical equilibrium prevailed and a balance between photo-ionization and electron dissociative recombination was maintained, but as $1/r^2$ outside the contact surface where the solar wind was mass-loaded and slowed down. However, the physical processes were different from the case reported in this paper where a contact surface has not formed and there is no photochemical equilibrium. The observed statistical $1/r$ behaviour should be interpreted with care. The Haser model assumes the plasma expands radially at constant speed, which means negligible effect of a solar wind electric field. There is indeed support for a strong quenching of this field from data from the RPC-ICA \citep{nilsson2015},\citep{nilsson2015b}, as they find the solar wind flow direction to be far away from radial in the near-nucleus environment. It is thus possible that we do have an almost complete quenching of the solar wind electric field at the close distances we investigate, also at this stage of comet activity where no diamagnetic cavity is formed. Noteworthy, however, from the large scatter around the fitted curve it is also possible that the observed statistical $1/r$ decay is accidental, being an average resulting from a combination of transport and solar wind E-field effects. More data analysis and modelling work are needed}

The 12.4 h spin period of the comet is clearly seen in the measured plasma parameters, and in the neutral gas density, as a 6.2-hour modulation. \textcolor{black}{The density of the neutrals and the plasma correlates well and the ratio $N_i/N_n \sim 1-2 \cdot10^{-6}$ at 10 km cometocentric distance and at 3.1 AU.} The $N_i/N_n$ ratio increases linearly with cometocentric distance, since $N_n$ decreases as $1/r^2$ and $N_i$ as $1/r$. 

Assuming an average plasma density of 200 cm$^{-3}$ at a distance of 10 km flowing radially outward at 700 m s$^{-1}$, in the northern hemisphere, suggests an hemispheric plasma outflow of $\sim 10^{20}$ s$^{-1}$. Further out, the plasma outflow will increase as more and more neutral molecules are being ionized with time.

We can also report that we see no clear signatures of the solar wind interaction with the cometary coma, in terms of plasma boundaries or separate plasma regions being formed, during the sub-100 km bound orbits when beyond 3.1 AU.  Short-scale (minutes to hours) variations (temporal and/or spatial) are often seen in the cometary plasma but are likely associated with effects of charged dust or jet-like features from the comet, secondary electron emission or variable electric fields, rather than from boundaries formed from the solar wind-comet interaction.


%
%
%
%
%
%
%

\begin{acknowledgments}
\textcolor{black}{Rosetta is a European Space Agency (ESA) mission with contributions from its member states and the National Aeronautics and Space Administration (NASA). NJTE was funded by SNSB (DNR35/13) and VR (621-2013-4191). We acknowledge the staff of CDDP and IC for the use of AMDA and the RPC Quicklook database. Work at the University of Bern on ROSINA COPS was funded by the State of Bern, the Swiss National Science Foundation, and the European Space Agency PRODEX Program. We thank the OSIRIS team for making the comet shape model available. The data used in this paper will soon be made available on the ESA Planetary Science Archive and available upon request until that time.} 
\end{acknowledgments}

%

\end{thebibliography}

\end{article}
%
%
%
%
%
%
 
 \clearpage
 
 \clearpage

%
%


\end{document}